\documentclass[aps,superscriptaddress,twocolumn,showpacs,dvips]{revtex4}
\usepackage{feynmp}
\usepackage{amssymb}
\usepackage{epsfig}
\usepackage{graphicx}
\usepackage{subfigure}
\usepackage{hyperref}
\usepackage{bbm}
\usepackage{gensymb}
\usepackage{float}
\usepackage{color}

\begin{document}

\title{Novel behavior of upper critical field due to nematic order in $d$-wave
superconductors}

\author{Jing-Rong Wang}
\affiliation{Department of Modern Physics, University of Science and
Technology of China, Hefei, Anhui 230026, P. R. China}
\author{Guo-Zhu Liu}
\affiliation{Department of Modern Physics, University of Science and
Technology of China, Hefei, Anhui 230026, P. R. China}

\begin{abstract}
In recent years, there have been increasing experimental evidence
suggesting the existence of an electronic nematic state in a number
of unconventional $d$-wave superconductors. Interestingly, it is
expected that the long-range nematic order can coexist with $d$-wave
superconductivity. We analyze the in-plane upper critical field
$H_{c2}$ after taking the influence of nematic state into account,
and find that the four-fold oscillation of angle-dependent $H_{c2}$
in a pure $d$-wave superconducting state is turned into a novel
two-fold oscillation pattern by a weak nematic order. Moreover, such
effect is much more significant at higher temperatures. These
behaviors are measurable and may be used to probe the predicted
coexisting nematic order in $d$-wave superconductors. In addition,
we show that the concrete angular dependence of $H_{c2}$ and the
positions of its maximum can be strongly affected by several
parameters, including temperature and $T_c$.
\end{abstract}

\pacs{74.20.Rp; 74.25.Dw; 74.60.Ec}

\maketitle


In the past three decades, there have been a plenty of theoretical
and experimental studies of the unusual and elusive properties of
unconventional superconductors \cite{Stewart84, Lohneysen07,
Stockert12, Lee06, Stewart11}, among which the high-$T_c$ cuprate
\cite{Lee06} and heavy fermion \cite{Stewart84, Lohneysen07,
Stockert12} superconductors have attracted particular interest. It
is well-known that cuprates and heavy fermion compounds all have
layered structures, with the physical quantities defined along the
$c$-axis being quite different from those of the basal $a$-$b$
planes. Generically, the electronic properties turn out to be
homogeneous in the $a$-$b$ planes. However, the past decade has
witnessed an increasing number of experimental evidence
\cite{Ando02, Hinkov08, Daou10, Lawler10, Okazaki11,
Chuang10,Chu10,Song11} which suggests the existence of a strong
anisotropy in the electronic properties within the $a$-$b$ planes in
some of these superconductors. The earliest evidence of such
electronic anisotropy comes from the transport measurements
performed on YBa$_{2}$Cu$_{3}$O$_{6+\delta}$ \cite{Ando02}. The
subsequent experiments have uncovered more extensive evidences in
the same compound \cite{Hinkov08, Daou10} and also in
Bi$_{2}$Sr$_{2}$CaCu$_{2}$O$_{8+\delta}$ \cite{Lawler10}. Moreover,
a strong anisotropy is found in the so-called hidden-order phase of
heavy fermion compound URu$_2$Si$_2$ \cite{Okazaki11}. It is
interesting that iron-based superconductors have also joined the
growing family with strong anisotropy \cite{Chuang10,Chu10,Song11}.

The strong anisotropy observed in the unconventional superconductors
can hardly be explained by the small intrinsic lattice anisotropy.
In fact, it is universally attributed to the formation of a novel
electronic nematic order that spontaneously breaks the $C_4$
symmetry down to a $C_2$ symmetry \cite{Kivelson98, Kivelson03,
Fradkin, Fradkin2, Vojta09}. In recent years, the phase transition
that leads to such an electronic nematic order and the associated
nematic critical behaviors have stimulated intensive research
effort, especially in the context of cuprates \cite{Kivelson98,
Kivelson03, Fradkin, Fradkin2, Vojta09}. In particular, the
interaction between the critical fluctuation of nematic order
parameter and gapless fermions \cite{Vojta00, Kim08, Huh08, Xu08,
Fritz09, Wang11, Liu12, WangLiu, WangJRLiu} is found to induce a
series of anomalous behaviors.

\begin{figure}
\includegraphics[width=3.3in]{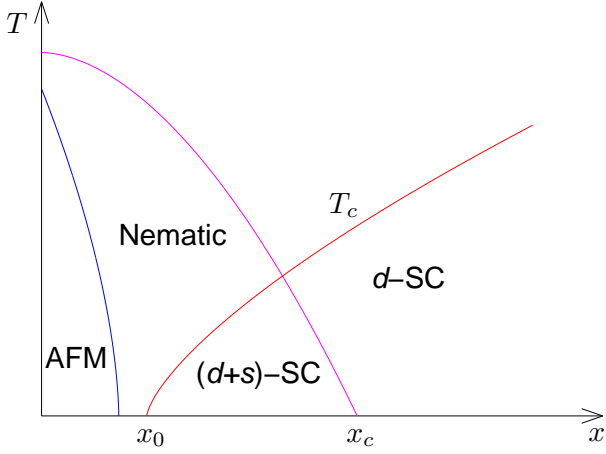}
\caption{Schematic phase diagram of unconventional superconductors.
$x_0$ denotes superconducting (SC) QCP and $x_c$ denotes nematic
QCP. The superconducting and nematic orders are supposed to coexist
in the region, dubbed (d+s)-SC, that is surrounded by $T_c$, $T_{n}$
and segment between $x_0$ and $x_c$.}\label{Fig:phase}
\end{figure}

In the meantime, much theoretical work has been done to address the
very interesting possibility that a nematic quantum phase transition
takes place in a $d$-wave superconductor \cite{Vojta00, Kim08,
Huh08, Xu08, Fritz09, Wang11, Liu12, WangLiu, WangJRLiu}. It was
found that the corresponding nematic quantum critical point (QCP),
dubbed $x_c$ on the schematic phase diagram Fig.~\ref{Fig:phase},
exhibits a series of nontrivial properties \cite{Vojta00, Kim08,
Huh08, Xu08, Fritz09, Wang11, Liu12, WangLiu, WangJRLiu}. While the
nematic QCP has been investigated extensively, little attention has
been paid to the broad region surrounded by $T_c$, $T_{n}$ and the
segment between $x_c$ and superconducting QCP $x_0$, where the
nematic and superconducting orders are supposed to coexist. If the
nematic-SC coexisting state really exists, it is interesting to
examine the influence of such state on observable quantities.

Amongst the observable quantities that can effectively reflects the
influence of nematic order, here we focus on the in-plane upper
critical field $H_{c2}$. Recently, in-plane $H_{c2}$ has played an
important role in the determination of the precise gap symmetry of
$d$-wave superconductors \cite{Won94, Takanaka95, Weickert06,
Vieyra11}. The main reason is that the concrete angular dependence
of in-plane $H_{c2}$ is intimately related to the angular dependence
of $d$-wave superconducting gap. The formation of an electronic
nematic order in a $d$-wave superconductor is always accompanied by
a change of the gap symmetry, hence the in-plane $H_{c2}$ turns out
to be a suitable quantity to characterize the physical effects
caused by nematic order.

As pointed out previously \cite{Vojta00}, the nematic order
developed in a $d_{x^2 - y^2}$-wave superconductor leads naturally
to a superconducting state with a mixed $d_{x^2 - y^2} + s$-wave
gap, where the additional $s$-wave component reflects the
modification of the superconducting order parameter in respond to
the nematic order. Due to this relationship, one should be able to
study the influence of the nematic order by examining how the
additional $s$-wave gap changes the angular dependence of in-plane
$H_{c2}$.

In this paper, we calculate in-plane $H_{c2}$ after taking into
account the effects of nematic order. On the basis of these results,
we will show that the well-known four-fold oscillation of
angle-dependent $H_{c2}$ in pure $d_{x^2 - y^2}$ superconductors is
strongly modified by the nematic order and completely turned into a
two-fold oscillation even when the additional $s$-wave gap induced
by nematic order is still quite weak. Moreover, the effects of
nematic order is more significant at higher temperatures. We also
study how $H_{c2}$ is affected by various physical parameters, such
as temperature, $T_c$ and fermion velocity, and find that the
concrete angular-dependence of $H_{c2}$ is very sensitive to these
parameters. We will perform calculations of $H_{c2}$ in the contexts
of both heavy fermion and cuprate superconductors, and then make a
comparison between the behaviors of $H_{c2}$ in these two types of
supercondcutors. Our predictions on $H_{c2}$ are unambiguous and
could be directly verified by experiments.

We consider a general layered $d$-wave superconductor. The physical
properties in the fundamental a-b plane is nearly isotropic, but the
coupling between a-b planes is relatively weaker. The Fermi surface
of layered superconductor usually has a complicated shape. To
simplify our calculations, it is convenient to choose a rippled
cylinder Fermi surface with a dispersion \cite{Thalmeier05,
Vorontsov07}
\begin{eqnarray}
\varepsilon(\mathbf{\mathbf{k}})=\frac{k_{x}^{2}+k_{y}^{2}}{2m} -
2t_{c}\cos(k_{z}c).
\end{eqnarray}
Now suppose an external magnetic field $H$ is imposed to the basal
plane. For type-II superconductors, any field $H$ weaker than the
lower critical field $H_{c1}$ cannot penetrate the superconductor
due to the Meissner effect. As $H$ exceeds $H_{c1}$ and further
grows, the superconducting gap is gradually destructed by the
orbital effect in many superconductors. The gap is entirely
suppressed once $H$ reaches an upper critical field $H_{c2}$, which
can be obtained by solving the corresponding linearized gap
equation. Under certain circumstances, the Pauli paramagnetic
limiting effect can also suppress the gap by breaking spin singlet
pairs, and may even be more important than the orbital effect in
some peculiar superconductors \cite{Weickert06, Vieyra11}. In order
to make a general analysis, we consider both of these two effects in
the following.

It it useful to rewrite the magnetic field vector $\mathbf{H}$ in
terms of a vector potential. We define the $a$-axis as
$x$-coordinate and the $b$-axis as $y$-coordinate, and choose a
vector potential
\begin{eqnarray}
\mathbf{A} = \left(0,0,H(-x\sin\theta+y\cos\theta)\right),
\end{eqnarray}
where $\theta$ is the angle between in-plane $H$ and a-axis. For
conventional $s$-wave superconductors, the paring gap is isotropic
and $H_{c2}$ is therefore $\theta$-independent. For $d$-wave
superconductors, however, the gap is strongly anisotropic, so
$H_{c2}$ becomes $\theta$-dependent. Now the field vector is of the
form
\begin{eqnarray}
\mathbf{H} &=& \mathbf{\nabla}\times\mathbf{A}
=\left(H\cos\theta,H\sin\theta,0\right).
\end{eqnarray}
The generalized derivative operator is given by
\begin{eqnarray}
\mathbf{\Pi}(\mathbf{R}) &=& -i\mathbf{\nabla}_{\mathbf{R}} +
2e\mathbf{A}(\mathbf{R}) \nonumber \\
&=& -i\partial_{x}\mathbf{e}_{x} + -i\partial_{y}\mathbf{e}_{y}
\nonumber \\
&& + \left(-i\partial_{z} + 2eH\left(-x\sin\theta +
y\cos\theta\right)\right)\mathbf{e}_{z}. \nonumber
\end{eqnarray}

Following the general methods presented in previous papers
\cite{Helfand66, Werthamer66, Scharnberg80, Lukyanchuk87,
Shimahara96, Suginishi06}, one can obtain the following linearized
gap equation:
\begin{eqnarray}
-\ln(\frac{T}{T_{c}})\Delta(\mathbf{R}) &=&
\int_{0}^{+\infty}d\eta\frac{\pi T}{\sinh(\pi T\eta)}
\int_{-\pi}^{\pi}\frac{d\chi}{2\pi}\int_{0}^{2\pi}\frac{d\varphi}{2\pi}
\nonumber \\
&& \times \gamma_{\alpha}^2(\hat{\mathbf{k}}) \left\{1-
\cos\left[\eta\left(h'+\frac{1}{2} \mathbf{v}_{F}(\hat{\mathbf{k}})
\right.\right.\right. \nonumber \\
&&\left.\left.\left.\cdot \mathbf{\Pi}(\mathbf{R})\right)\right]
\right\}\Delta(\mathbf{R}), \label{eqn:GapL}
\end{eqnarray}
where $\chi = k_{z}c$ and the function $\Delta(\mathbf{R})$ is
\begin{eqnarray}
\Delta(\mathbf{R}) = \left(\frac{2eH}{\pi}\right)^{\frac{1}{4}}
e^{-eH\left(x\sin\theta-y\cos\theta\right)^{2}}.
\end{eqnarray}
Here, for simplicity, we do not include the Landau level mixing
\cite{Weickert06,Vieyra11,Lukyanchuk87},which will not affect our
conclusion. For the chosen cylinder Fermi
surface, the Fermi velocity vector takes the form \cite{Thalmeier05}
\begin{eqnarray}
\mathbf{v}_{F}(\hat{\mathbf{k}})=v_{a}\cos\varphi \mathbf{e}_{x}
+v_{a}\sin\varphi \mathbf{e}_{y}+v_{c}\sin\chi \mathbf{e}_{z}.
\label{eqn:FermionV}
\end{eqnarray}
The Fermi velocity component along $c$-axis is $v_{c} = 2t_{c}c$.
The velocity $v_{a}$ is defined as $v_{a}
=v_{0}\sqrt{1+\lambda\cos(\chi)}$, where $v_{0} = \frac{k_{F0}}{m}$
with Fermi momentum being related to Fermi energy $\epsilon_{F}$ by
$k_{F0} = \sqrt{2m\epsilon_{F}}$. The ratio $\frac{v_{c}}{v_{0}} =
\lambda\gamma$ with $\lambda = \frac{2t_{c}}{\epsilon_{F}}$ and
$\gamma = \frac{ck_{F0}}{2}$. Moreover, we define $h' =
-\frac{g\mu_{B}H}{2}$, where $\mu_B$ is Bohr magneton and $g$ is the
gyromagnetic ratio. The orbital effect of magnetic field is
reflected in the factor $\mathbf{v}_{F}(\mathbf{k}) \cdot
\Pi(\mathbf{R})$, whereas the Pauli paramagnetic effect is reflected
in the factor $h'$. The concrete behavior of $H_{c2}$ is determined
by the complex interplay of these two effects.

\begin{figure}[htbp]
\center
\includegraphics[width=3.2in]{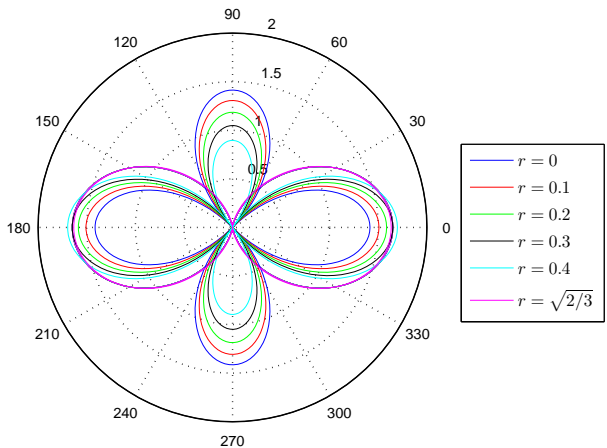}
\caption{Magnitude of $d+s$-wave gap for different values of $r$.}
\label{Fig:GapShape}
\end{figure}

We choose the direction of field $H$ as a new $z'$ axis, and then
define
\begin{eqnarray}
\left(\mathbf{e}_{x}',\mathbf{e}_{y}',\mathbf{e}_{z}'\right) =
(\mathbf{e}_{x}\sin\theta - \mathbf{e}_{y}\cos\theta,
-\mathbf{e}_{z}, \mathbf{e}_{x}\cos\theta +
\mathbf{e}_{y}\sin\theta). \nonumber
\end{eqnarray}
In the coordinate frame spanned by $(\mathbf{e}_{x}',
\mathbf{e}_{y}', \mathbf{e}_{z}')$, we have
\begin{eqnarray}
\mathbf{v}_{F}(\hat{\mathbf{k}}) &=& v_{a}\sin(\theta-\varphi)
\mathbf{e}_{x}' - v_{c}\sin(\chi)\mathbf{e}_{y}' \nonumber \\
&& + v_{a}\cos(\theta-\varphi)\mathbf{e}_{z}',\label{eqn:FermionV2}
\end{eqnarray}
and
\begin{eqnarray}
\mathbf{\Pi}(\mathbf{R}) &=& \sqrt{eH}
\left[\left(a_{+}+a_{-}\right)\mathbf{e}_{x}' -
i\left(a_{+}-a_{-}\right)\mathbf{e}_{y}' \right.\nonumber
\\
&&\left.+\sqrt{2}a_{0}\mathbf{e}_{z}'\right], \label{eqn:PiDef}
\end{eqnarray}
where
\begin{eqnarray}
a_{\pm} &=& \frac{1}{2\sqrt{eH}}\left[-i\sin\theta\partial_{x} +
i\cos\theta\partial_{y} \mp \partial_{z}\right. \nonumber \\
&&\left.\pm 2ieH(x\sin\theta-y\cos\theta)\right], \\
a_{0} &=& \frac{1}{\sqrt{2eH}}\left[-i\partial_{x}\cos\theta -
i\partial_{y}\sin\theta \right],
\end{eqnarray}
which satisfy $[a_{-},a_{+}] = 1$ and $[a_{\pm},a_{0}] = 0$.

In Eq.(\ref{eqn:GapL}), the gap symmetry is encoded in the function
$\gamma_{\alpha}(\hat{\mathbf{k}})$. For isotropic $s$-wave pairing,
$\gamma_{s}(\hat{\mathbf{k}}) = 1$. For $d_{x^2 - y^2}$-wave
pairing, one can write the gap as \cite{Suginishi06}
\begin{eqnarray}
\gamma_{d}(\hat{\mathbf{k}}) = \sqrt{2}\cos(2\varphi)
\end{eqnarray}
To make a general analysis, we consider a mixed gap
\begin{eqnarray}
\gamma_{d+s}(\hat{\mathbf{k}}) = \sqrt{1-r^2}\sqrt{2}\cos(2\varphi)
+ r,\label{eqn:Gammads}
\end{eqnarray}
where $r$ is a tuning parameter. When $r=0$, the above mixed gap
reduces to the pure $d_{x^2 - y^2}$-wave gap, whereas $r =
\sqrt{2/3}$ corresponds to the upper limit of the influence of
nematic order. For $r \subset (0,\sqrt{2/3})$, there is a mixing
between $d_{x^2 - y^2}$-wave and $s$-wave gaps, and there are still
four gapless nodes. Apparently, $r$ measures the ratio of $s$-wave
component to the mixed gap. Once a nematic order is formed in a
$d_{x^2 - y^2}$-wave superconductor, the effective superconducting
gap is described by Eq.(\ref{eqn:Gammads}), with $r$ reflecting the
influence of the nematic order.

In order to make the phenomenon of gap mixing more transparent, we
plot the mixed gap in Fig.~\ref{Fig:GapShape}. When $r=0$, the pure $d_{x^2 -
y^2}$-wave gap exhibits a discrete $C_4$ symmetry. Once $r$ becomes
finite, the $C_4$ symmetry is broken down to a $C_2$ symmetry, which
is obviously driven by the formation of nematic order. It is
interesting to examine how $r$, and other physical parameters as
well, influence the angular dependence of $H_{c2}$. At first glance,
it seems that $H_{c2}$ should always exhibit its maximum along the
antinodal direction ($\theta = 0, \pi$) since the mixed gap is
maximal in this direction. As will be shown below, this is actually
not the case. For certain parameters, $H_{c2}$ could exhibit its
maximum along the nodal direction where the mixed gap vanishes. In
addition, although the mixed gap has four nodes for $r <\sqrt{2/3}$,
the four-fold oscillation pattern of $H_{c2}$ can be completely
changed by the nematic order.

Substituting Eq.~(\ref{eqn:FermionV2}), Eq.~(\ref{eqn:PiDef}) and
Eq.~(\ref{eqn:Gammads}) into the linearized gap equation
(\ref{eqn:GapL}) and making average over the ground state
$\Delta(\mathbf{R})$, following the procedure presented in
Ref.~\cite{Helfand66}, we finally obtain an integral equation of
$H_{c2}$,
\begin{eqnarray}
\ln \frac{1}{t} &=& \int_{0}^{+\infty}du \frac{1-\cos(hu)}{\sinh(u)}
\left\{\int_{-\pi}^{\pi}\frac{d\chi}{2\pi}
\int_{0}^{2\pi}\frac{d\varphi}{2\pi}\right.\nonumber \\
&&\times
\left[(1-r^2)\left(1+\cos(4\varphi)\cos(4\theta)\right)\right.
\nonumber \\
&& +
\left.2\sqrt{2}r\sqrt{1-r^2}\cos(2\varphi)\cos(2\theta)+r^2\right]
\nonumber \\
&& \left.\times\exp\left[-\rho u^2 \left[\sin^{2}(\varphi) +
\frac{v_{c}^2}{v_{a}^2}\sin^{2}(\chi) \right]\right] \right\},
\nonumber \\ \label{eq:Hc2FinalExp}
\end{eqnarray}
where
\begin{eqnarray}
t = \frac{T}{T_{c}}, \quad h = \frac{g\mu_{B}}{2\pi k_{B}T}eH_{c2},
\quad \rho = \frac{\hbar v_{a}^{2}}{8\pi^2 k_{B}^{2}T^2}eH_{c2}.
\end{eqnarray}
The angular dependence of in-plane $H_{c2}$ can be obtained by
solving this equation.

The equation of $H_{c2}$ contains a number of parameters, and cannot
be really solved without choosing concrete magnitudes for these
parameters. Although our analysis is actually material independent,
we need to consider some realistic samples so as to obtain a
detailed angular dependence of $H_{c2}$. In this paper we consider
two types of $d$-wave superconductors: cuprates and heavy fermion
compounds. It is fair to say that the experimental evidence for
nematic order in cuprates is far more extensive and compelling than
that in heavy fermion compounds. To date, there are only limited
reports suggesting the existence of nematic state in one single
heavy fermion compounds. However, it is still useful to make a
general analysis and compare the unusual behaviors of $H_{c2}$
obtained in different sorts of compounds. As will be shown below,
this can help us to gain valuable insights on the difference between
orbital effect and Pauli paramagnetic effect. Moreover, $H_{c2}$
manifests unexpected properties even in the absence of nematic
order, which makes it not only interesting but necessary to perform
a detailed analysis in the contexts of heavy fermion compounds.

\begin{figure}[htbp]
\center
\includegraphics[width=3.1in]{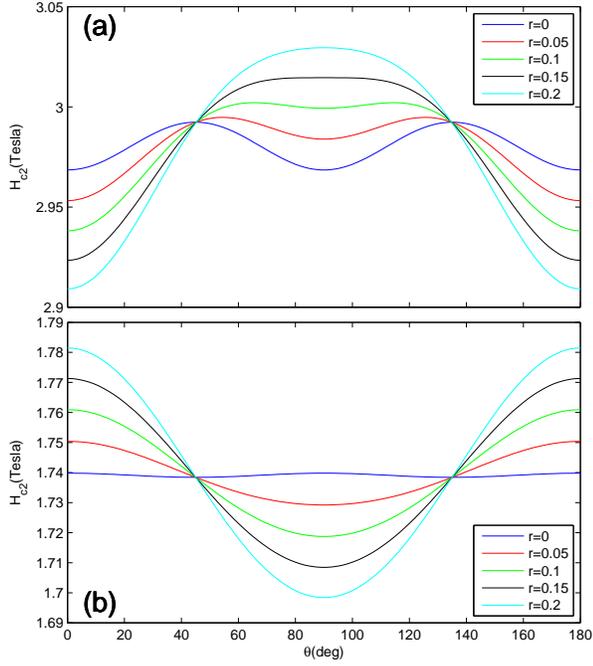}
\caption{The $\theta$-dependence of $H_{c2}$ for different values of
$r$. Here, $g=1$, $v_{0} = 2000\mathrm{m}/\mathrm{s}$, $\lambda =
0.5$ and $\gamma=1$. We suppose $T_{c} = 1K$, which is a typical
value for heavy fermion superconductors. (a) $t=0.1$; (b) $t=0.9$.}
\label{Fig:Hc2ThetaA}
\end{figure}

Fig.~\ref{Fig:Hc2ThetaA} shows the $\theta$-dependence of $H_{c2}$
for a number of different values of $r$. We choose $T_{c} = 1K$,
$v_{0} = 2000 \mathrm{m}/\mathrm{s}$ and $g=1$, which are suitable
quantities for heavy fermion compounds. The results are obtained at
two representative temperatures, $t = 0.1$ and $t = 0.9$,
respectively. Though heavy fermion compounds have layered
structures, the coupling between different layers are not very
small, so we choose $\lambda = 0.5$. In addition, we take $\gamma=1$
throughout the whole paper.

Let us first concentrate on Fig.~\ref{Fig:Hc2ThetaA}(a). We can see
that the $\theta$-dependence of $H_{c2}$ exhibits a four-fold
oscillation at $r=0$, which is well consistent with previous results
\cite{Won94,Takanaka95}. As $\theta$ changes from $0$ to $\pi$,
there are two peaks for $H_{c2}$, locating at $\pi/4$ and $3\pi/4$
respectively. As $r$ is growing continuously, these two peaks move
away from their original positions and approach each other. Once $r$
exceeds certain critical value $r_c$, they converge to one single
peak located at $\theta = \pi/2$. As shown in
Fig.~\ref{Fig:Hc2ThetaA}(a), $H_{c2}$ exhibits a two-fold
oscillation at $r = 0.2$, reflecting the strong influence of the
nematic order. After careful calculations, we find the critical
value $r_c \approx 0.149$ for $t = 0.1$. For $r$ greater than $r_c$
(remaining smaller than $\sqrt{2/3}$), the mixed gap still has four
nodes as $\theta$ grows from $0$ to $2\pi$, but $H_{c2}$ exhibits
only one peak from $0$ to $\pi$. Apparently, a small amount of
nematic order-induced $s$-wave gap is strong enough to trigger the
four-to-two fold transition of $H_{c2}$.

\begin{figure}[htbp]
\center
\includegraphics[width=3.1in]{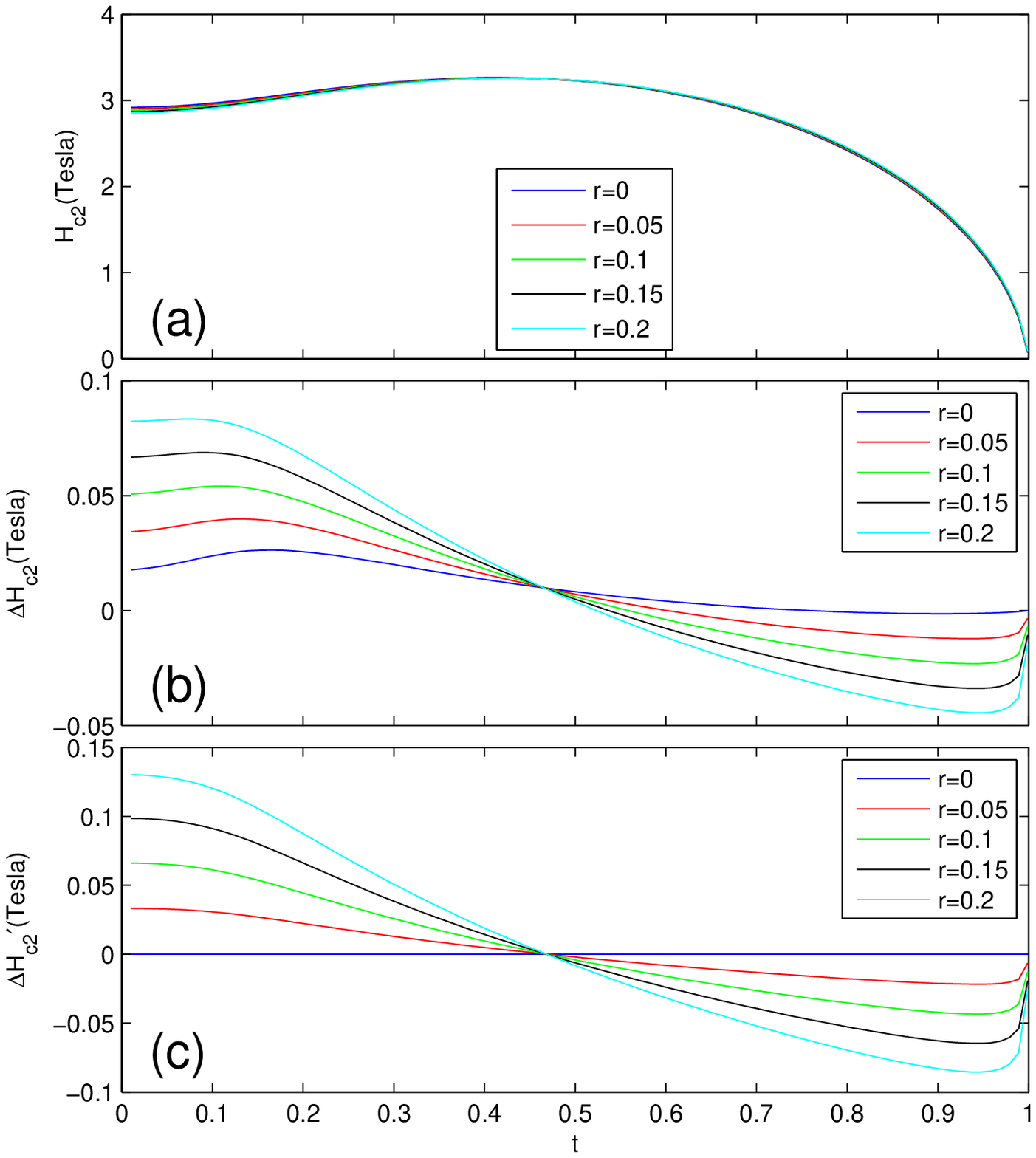}
\caption{Relation between $\Delta H_{c2}$, $\Delta H_{c2}'$ and $t$
with $g=1$, $v_{0}=2000m/s$, $\lambda=0.5$, $\gamma=1$, $T_{c}=1K$
for different values of $r$. (a) $H_{c2}$ for $\theta=0^{\degree}$;
(b) $\Delta H_{c2} = H_{c2}(\theta=45^{\degree}) -
H_{c2}(\theta=0^{\degree})$; (c) $\Delta
H_{c2}'=H_{c2}(\theta=90^{\degree})-H_{c2}(\theta=0^{\degree})$.}
\label{Fig:Hc2tA}
\end{figure}

Similar to Fig.~\ref{Fig:Hc2ThetaA}(a), $H_{c2}$ presented in
Fig.~\ref{Fig:Hc2ThetaA}(b) also exhibits a four-fold oscillation at
small $r$ and a two-fold oscillation when $r$ is greater than a
critical value $r_c$. We find the critical value $r_c \approx 0.013$
for $t = 0.9$. This result indicates that the effect of nematic
order is much more significant at high temperatures.

In realistic experiments, the existence of nematic state is usually
identified by measuring the spatial dependence of certain observable
quantities\cite{Fradkin,Fradkin2}, such as specific heat
and thermal conductivity. The in-plane $H_{c2}$ provides a new route
to probe the nematic state. One advantage of this route is that the
nematic state can be determined by measuring the angular dependence
of $H_{c2}$ even when the nematic order is rather weak, because the
two-fold oscillation can be easily distinguished from the four-fold
oscillation. Another interesting result is that the critical value
$r_c$ obtained at high temperatures is much smaller than that of low
temperatures. If $r$ is fixed at a given value, such as $r = 0.05$,
there should be a four-to-two fold transition of $H_{c2}$ as $T$
increases from $T=0$ to $T_c$.

One can notice a very important difference between
Fig.~\ref{Fig:Hc2ThetaA}(a) and Fig.~\ref{Fig:Hc2ThetaA}(b):
$H_{c2}(\theta)$ exhibits a minimum at $\theta = 0$ for $t = 0.1$,
but exhibits a maximum at $\theta = 0$ for $t = 0.9$. Two main
conclusions can be drawn from these results. First, the maximum of
$H_{c2}$ is not necessarily along the anti-nodal directions where
the $d_{x^2 - y^2}$-wave gap is maximal. Second, $H_{c2}$ exhibits
apparently different $\theta$-dependences at low and high
temperatures, and there is a transition of the oscillation pattern
of $H_{c2}$ at certain critical temperature $T_{\theta}$.

\begin{figure}[htbp]
\center
\includegraphics[width=3.1in]{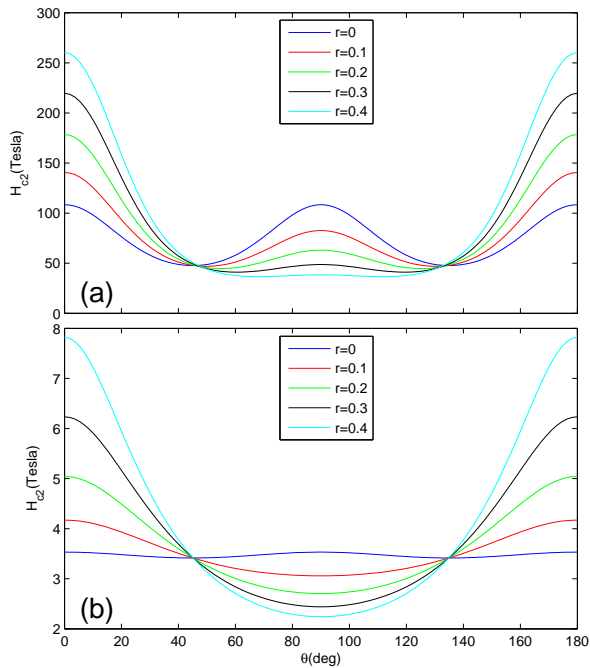}
\caption{$\theta$-dependence of $H_{c2}$ for different value of $r$.
Here, $g=0.1$, $v_{0}=10^{5}\mathrm{m}/\mathrm{s}$, $\lambda =
0.01$, and $\gamma=1$. We suppose $T_{c} = 40\mathrm{K}$, which is a
suitable value for cuprates. (a) $t=0.1$; (b) $t=0.95$.}
\label{Fig:Hc2ThetaB}
\end{figure}

We feel it necessary to examine the effects of different
temperatures on the behaviors of $H_{c2}$ in more details. We show
the $t$-dependence of $H_{c2}(\theta = 0^{\degree})$ in
Fig.~\ref{Fig:Hc2tA}(a), that of $\Delta H_{c2} = H_{c2}(\theta =
45^{\degree}) - H_{c2}(\theta=0^{\degree})$ in
Fig.~\ref{Fig:Hc2tA}(b) and that of $\Delta H_{c2}' = H_{c2}(\theta
= 90^{\degree}) - H_{c2}(\theta=0^{\degree})$ in
Fig.~\ref{Fig:Hc2tA}(c). First of all, notice that $H_{c2}$ shown in
Fig.~\ref{Fig:Hc2tA}(a) is not monotonic at lower temperatures,
which agrees well with the results of Ref.~\cite{Vieyra11}. Such
non-monotonicity is presumably due to the Pauli paramagnetic effect
\cite{Vieyra11}. Furthermore, notice that the differences $\Delta
H_{c2}$ and $\Delta H_{c2}'$, shown in Fig.~\ref{Fig:Hc2tA}(b) and
Fig.~\ref{Fig:Hc2tA}(c) respectively, are positive at lower
temperatures but negative at higher temperatures. This property
clearly demonstrates the existence of a temperature-driven
transition between two different oscillation patterns of
$\theta$-dependent $H_{c2}$.

\begin{figure}[htbp]
\center
\includegraphics[width=3.1in]{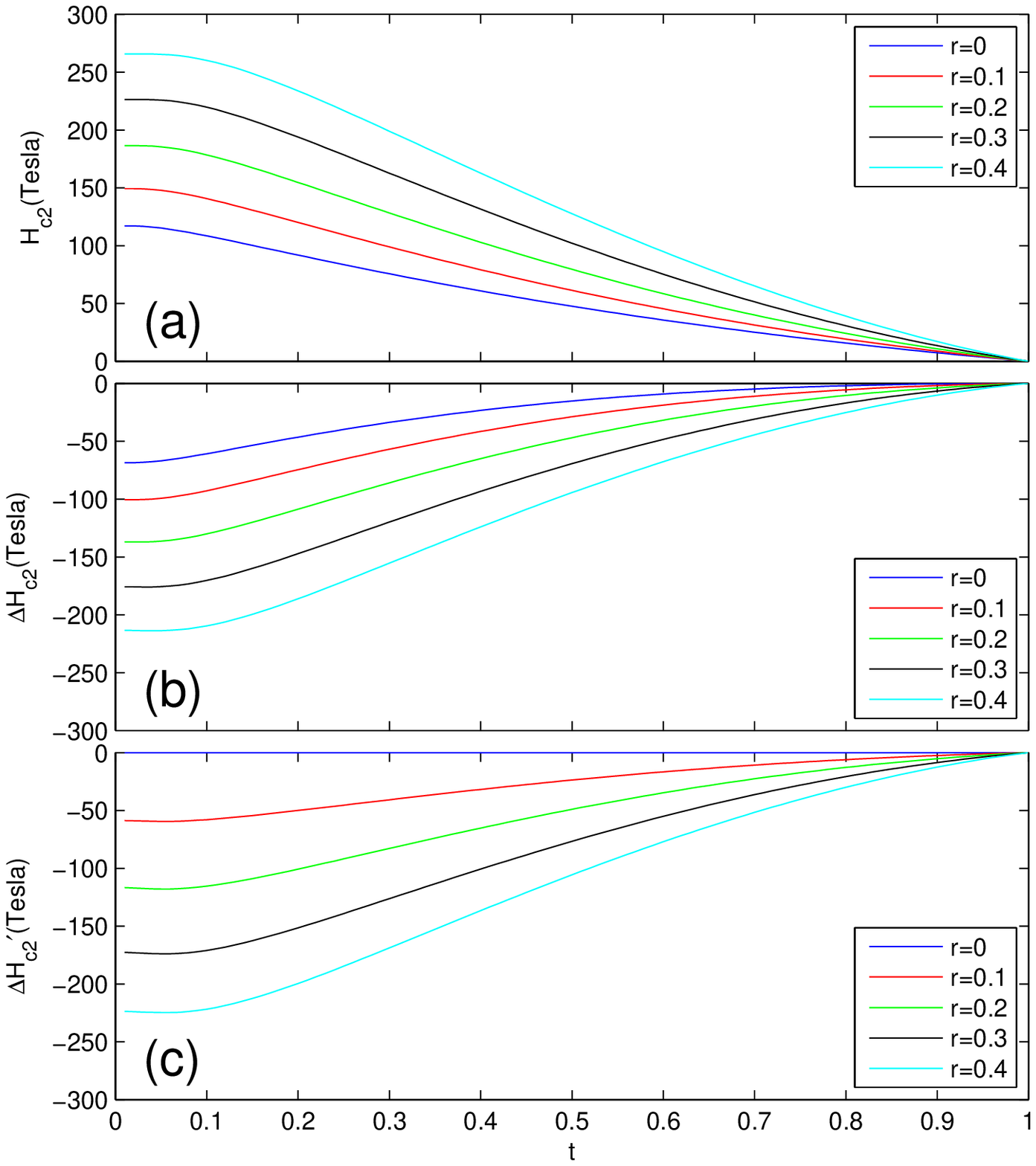}
\caption{Relation between $\Delta H_{c2}$, $\Delta H_{c2}'$ and $t$
with $g=0.1$, $v_{0}=10^{5}m/s$, $\lambda=0.01$, $\gamma=1$, $T_{c}
= 40K$ for different values of $r$. (a) $H_{c2}$ for $\theta =
0^{\degree}$; (b) $\Delta H_{c2} = H_{c2}(\theta=45^{\degree}) -
H_{c2}(\theta=0^{\degree})$; (c) $\Delta H_{c2}' =
H_{c2}(\theta=90^{\degree}) - H_{c2}(\theta=0^{\degree})$.}
\label{Fig:Hc2tB}
\end{figure}

The above theoretical analysis can be readily applied to study the
behaviors of $H_{c2}$ in the contexts of cuprates. We show the
$\theta$-dependence of $H_{c2}$ in Fig.~\ref{Fig:Hc2ThetaB} for
several values of $r$ after assuming $T_c = 40\mathrm{K}$,
$v_{0}=10^5m/s$ and $g = 0.1$, which are typical values for
cuprates. Since the coupling between different layers in cuprates is
quite small, we take $\lambda = 0.01$. Fig.~\ref{Fig:Hc2ThetaB}(a)
is obtained at $t = 0.1$, and Fig.~\ref{Fig:Hc2ThetaB}(b) at $t =
0.95$. It is easy to see that $H_{c2}$ exhibits a four-fold
oscillation at $r = 0$ and a two-fold oscillation once $r$ becomes
sufficiently large. This feature is qualitatively the same as that
in the case of heavy fermion compounds. Once again, the formation of
a nematic order is responsible for this four- to two-fold
oscillation transition.

However, $H_{c2}$ shown in Fig.~\ref{Fig:Hc2ThetaB} differs from
that in Fig.~\ref{Fig:Hc2ThetaA} in several aspects. First, it is
widely believed that the Pauli paramagnetic effect plays little role
in cuprates. We have re-calculated $H_{c2}$ under the same
parameters given in the last paragraph after removing the Pauli
effect, and found a nearly negligible change of the behavior of
$H_{c2}$. This indicates that the Pauli paramagnetic effect can be
guaranteed to be unimportant once an appropriate set of parameters
are chosen. Second, as shown in Fig.~\ref{Fig:Hc2ThetaB}(a), the
maximal value of $H_{c2}$ is well beyond $100$ Tesla at low
temperature $t = 0.1$. This result is actually expectable since it
has been known for a long time that $H_{c2}$ of cuprates is roughly
$100 \sim 200$ Tesla. Such a strong magnetic field is apparently too
large to be realized in laboratories, so the predicted
low-temperature behavior of $H_{c2}$ can hardly be tested by
experiments. Moreover, we find the critical value $r_{c} = 0.513$
for $t = 0.1$, which means the four-fold oscillation of $H_{c2}$
will not be converted to two-fold as long as the nematic order is
not extremely strong.

We next consider the results obtained at some finite temperature,
shown in Fig.~\ref{Fig:Hc2ThetaB}(b). If the temperature is fixed at
$t = 0.95$, the magnitude of $H_{c2}$ is always smaller than $10$
Tesla, which is not very strong and can be achieved in laboratories.
In addition, we find the critical value of $r$ at $t = 0.95$ is
quite small: $r_c = 0.043$. Therefore, the effects of nematic order
on $H_{c2}$ is much more significant at higher temperatures, and can
be most easily measured at temperatures near $T_c$.

Similar to the case of heavy fermion compounds, there might be a
temperature-driven transition from four-fold oscillation of $H_{c2}$
to two-fold oscillation if $r$ is fixed at a suitable value, such as
$r = 0.1$. However, from Fig.~\ref{Fig:Hc2ThetaB}, we see that
$H_{c2}$ always displays its maximum at $\theta = 0$ and $\theta =
\pi$ in the whole range of $0 \le t < 1$. This property is quite
different from that found in the case of heavy fermion compounds. We
attribute this difference to the Pauli paramagnetic effect, which
plays no role in cuprates but could be very important in heavy
fermion compounds.

It is helpful to plot the $t$-dependence of $H_{c2}(\theta =
0^{\degree})$ in Fig.~\ref{Fig:Hc2tB}(a), that of $\Delta H_{c2} =
H_{c2}(\theta = 45^{\degree}) - H_{c2}(\theta=0^{\degree})$ in
Fig.~\ref{Fig:Hc2tB}(b) and that of $\Delta H_{c2}' = H_{c2}(\theta
= 90^{\degree}) - H_{c2}(\theta=0^{\degree})$ in
Fig.~\ref{Fig:Hc2tB}(c). $H_{c2}$ shown in Fig.~\ref{Fig:Hc2tB}(a)
is monotonic as temperature is varying, which is understandable
since the Pauli effect is believed to be negligible. The differences
$\Delta H_{c2}$ and $\Delta H_{c2}'$, shown in
Fig.~\ref{Fig:Hc2tB}(b) and Fig.~\ref{Fig:Hc2tB}(c) respectively,
are always negative in the whole temperature range, which are in
sharp contrast with those shown in Fig.~\ref{Fig:Hc2tA}(b) and
Fig.~\ref{Fig:Hc2tA}(c).

In summary, we have performed a detailed analysis of the in-plane
upper critical field $H_{c2}$ in layered $d_{x^2 - y^2}$-wave
superconductors after including the influence of a long-range
nematic order that is supposed to coexist with superconductivity. On
the basis of our results, we have reached two main conclusions.
First, the well-known four-fold oscillation of the angle-dependent
$H_{c2}$ in pure $d$-wave superconductors can be turned into a novel
two-fold oscillation by a weak nematic order. The strength of the
nematic order required to trigger such a four-to-two fold transition
decreases rapidly as temperature grows, thus it is much easier to
probe the effects of nematic order and seek the predicted two-fold
oscillation of $H_{c2}$ at higher temperatures. Moreover, there may
be a four-to-two fold transition of $H_{c2}$ driven by temperature
at certain given strength of nematic order. Second, the concrete
angle-dependence of $H_{c2}$ is determined by the specific values of
a number of physical parameters including $T_{c}$, $t$, $g$,
$v_{0}$, $\lambda$ and $\gamma$. In particular, the positions of
maximum of $H_{c2}$ may rotate by $\pi/4$ or $\pi/2$ as temperature
increases from zero to $T_c$ in heavy fermion compounds. However, no
evidence of such behavior is observed in cuprates. Such a difference
is presumably due to the difference in the role played by Pauli
paramagnetic effect in these two kinds of superconductors.

We believe that the predicted two-fold oscillation of in-plane
$H_{c2}$ driven by nematic order is particularly interesting. This
novel oscillation pattern is unambiguous and could be clearly
identified in experiments provided it really occurs. More
importantly, such two-fold oscillation can happen even when the
nematic order is quite weak and the spatial anisotropy of other
physical quantities is not significant enough to be observed.
Therefore, this behavior may serve as an experimental signature for
the existence of a long-range nematic order in $d_{x^2 - y^2}$-wave
superconductors.

Another interesting result is that the angle-dependence of $H_{c2}$
is actually very sensitive to a number of physical parameters. A
more extensive and systematical analysis is required to fully reveal
the influence of all the relevant parameters on $H_{c2}$. Such
analysis is necessary since the concrete angle-dependence of
$H_{c2}$ proves to be a powerful tool in the determination of
precise gap symmetry of $d$-wave superconductors \cite{Won94,
Takanaka95, Weickert06, Vieyra11}. This issue will be addressed in
more details in a separate work.

In the above analysis, we have considered solely the superconductors
with a $d_{x^2 - y^2}$-wave gap symmetry. It is proposed that some
superconductors might have a $d_{xy}$ gap \cite{Vieyra11, Izawa02}.
We have applied the same formalism to the superconductors with a
$d_{xy}$-wave gap, and found that the effects of nematic order on
in-plane $H_{c2}$ are very similar to those in $d_{x^2 - y^2}$-wave
superconductors. There is only one difference: the maximum of
$H_{c2}$ rotates by $\pi/4$ compared to the case of $d_{x^2 -
y^2}$-wave gap.

J.R.W. is grateful to the support by the joint doctoral promotion
programme sponsored by Max-Planck-Gesellschaft and Chinese Academy
of Sciences. G.Z.L. is supported by the National Natural Science
Foundation of China under grants No. 11074234 and No. 11274286.


\begin{thebibliography}{99}

\bibitem{Stewart84}
G. R. Stewart, Rev. Mod. Phys. {\bf 56}, 755 (1984).

\bibitem{Lohneysen07}
H. v. L\"{o}hneysen, A. Rosch, M. Vojta, and P. W\"{o}lfle, Rev.
Mod. Phys. {\bf 79}, 1105 (2007).

\bibitem{Stockert12}
O. Stockert, S. Kirchner, F. Steglich, and Q. Si, J. Phys. Soc. Jpn.
{\bf 81}, 011001 (2011).

\bibitem{Lee06}
P. A. Lee, N. Nagaosa, and X.-G. Wen, Rev. Mod. Phys. {\bf 78}, 17
(2006).


\bibitem{Stewart11}
G. R. Stewart, Rev. Mod. Phys. {\bf 83}, 1589 (2011); A. Chubukov,
Annu. Rev. Condens. Matter Phys. {\bf 3}, 57 (2012).

\bibitem{Ando02}
Y. Ando, K. Segawa, S. Komiya, and A. N. Lavrov, Phys. Rev. Lett.
{\bf 88}, 137005 (2002).

\bibitem{Hinkov08}
V. Hinkov, D. Haug, B. Fauqu\'{e}, P. Bourges, Y. Sidis, A. Ivanov,
C. Bernhard, C. T. Lin, and B. Keimer, Science {\bf 319}, 597
(2008).

\bibitem{Daou10}
R. Daou, J. Chang, D. LeBoeuf, O. Cyr-Choini\`{e}re, F.
Lalibert\'{e}, N. Doiron-Leyraud, B. J. Ramshaw, R. Liang, D. A.
Bonn, W. N. Hardy, and L. Taillefer, Nature (London) {\bf 463}, 519
(2010).

\bibitem{Lawler10}
M. J. Lawler, K. Fujita, J. Lee, A. R. Schmidt, Y. Kohsaka, C. K.
Kim, H. Eisaki, S. Uchida, J. C. Davis, J. P. Sethna, and E.-A. Kim,
Nature {\bf 466}, 347 (2010).

\bibitem{Okazaki11}
R. Okazaki, T. Shibauchi,  H. J. Shi, Y. Haga, T. D. Matsuda, E.
Yamamoto, Y. Onuki, H. Ikeda, and Y. Matsuda, Science {\bf 331}, 439
(2011).

\bibitem{Chuang10}
T.-M. Chuang, M. P. Allan, J. Lee, Y. Xie, N. Ni, S. L.
Bud$^{\prime}$ko, G. S. Boebinger, P. C. Canfield, and J. C. Davis,
Science {\bf 327}, 181 (2010).

\bibitem{Chu10}
J.-H. Chu, J. G. Analytis, K. D. Greve, P. L. McMahon, Z. Islam, Y. Yamamoto, and I. R. Fisher,
Science {\bf 329}, 824 (2010).

\bibitem{Song11}
C.-L. Song, Y.-L. Wang, P. Cheng, Y.-P. Jiang, W. Li, T. Zhang,
Z. Li, K. He, L. Wang, J-F. Jia, H.-H. Hung, C. Wu, X. Ma, X. Chen,
and Q.-K. Xue, Science {\bf 332}, 1410 (2011).


\bibitem{Kivelson98}
S. A. Kivelson, E. Fradkin, and V. J. Emery, Nature (London) {\bf
393}, 550 (1998).

\bibitem{Kivelson03}
S. A. Kivelson, I. P. Bindloss, E. Fradkin, V. Oganesyan, J. M.
Tranquada, A. Kapitulnik and C. Howald, Rev. Mod. Phys. {\bf 75},
1201 (2003).

\bibitem{Fradkin}
E. Fradkin, S. A. Kivelson, M. J. Lawler, J. P. Eisenstein and A. P.
Mackenzie, Annu. Rev. Condens. Matter Phys. {\bf 1}, 153 (2010).

\bibitem{Fradkin2}
E. Fradkin, in \emph{Modern} \emph{Theories} \emph{of}
\emph{Many}-\emph{Particle} \emph{Systems} \emph{in}
\emph{Condensed} \emph{Matter} \emph{Physics}, edited by D. C.
Cabra, A. Honecker, and P. Pujol, Lecture Notes in Physics, Vol. 843
(Springer-Verlag, Berlin Heidelberg, 2012).

\bibitem{Vojta09}
M. Vojta, Adv. Phys. {\bf 58}, 699 (2009).

\bibitem{Vojta00}
M. Vojta, Y. Zhang, and S. Sachdev, Phys. Rev. B {\bf 62}, 6721
(2000); Int. J. Mod. Phys. B {\bf 14}, 3719 (2000).

\bibitem{Kim08}
E.-A. Kim, M. J. Lawler, P. Oreto, S. Sachdev, E. Fradkin, and S. A.
Kivelson, Phys. Rev. B {\bf 77}, 184514 (2008).

\bibitem{Huh08}
Y. Huh and S. Sachdev, Phys. Rev. B {\bf 78}, 064512 (2008).

\bibitem{Xu08}
C. Xu, Y. Qi, and S. Sachdev, Phys. Rev. B {\bf 78}, 134507 (2008).

\bibitem{Fritz09}
L. Fritz and S. Sachdev, Phys. Rev. B {\bf 80}, 144503 (2009).

\bibitem{Wang11}
J. Wang, G.-Z. Liu, and H. Kleinert, Phys. Rev. B {\bf 83}, 214503
(2011).

\bibitem{Liu12}
G.-Z. Liu, J.-R. Wang, and J. Wang, Phys. Rev. B {\bf 85}, 174525
(2012).

\bibitem{WangJRLiu}
J.-R. Wang and G.-Z. Liu, New J. Phys. {\bf 15}, 063007 (2013).

\bibitem{WangLiu}
J. Wang and G.-Z. Liu, New J. Phys. {\bf 15}, 073039 (2013).

\bibitem{Won94}
H. Won and K. Maki, Physica B {\bf 199}, 353 (1994).

\bibitem{Takanaka95}
K. Takanaka and K. Kuboya, Phys. Rev. Lett. {\bf 75}, 323 (1995).

\bibitem{Weickert06}
F. Weickert, P. Gegenwart, H. Won, D. Park, and K. Maki, Phys. Rev.
B {\bf 74}, 134511 (2006).

\bibitem{Vieyra11}
H. A. Vieyra, N. Oeschler, S. Seiro, H. S. Jeevan, C. Geibel, D.
Parker, and F. Steglich, Phys. Rev. Lett. {\bf 106}, 207001 (2011).

\bibitem{Thalmeier05}
P. Thalmeier, T. Watanabe, K. Izawa, and Y. Matsuda, Phys. Rev. B
{\bf 72}, 024539 (2005).

\bibitem{Vorontsov07}
A. B. Vorontsov and I. Vekhter, Phys. Rev. B {\bf 75}, 224501
(2007); \emph{ibid} {\bf 75}, 224502 (2007).

\bibitem{Helfand66}
E. Helfand, N. R. Werthamer, Phys. Rev. {\bf 147}, 288 (1966).

\bibitem{Werthamer66}
N. R. Werthamer, E. Helfand, and P. C. Hohenberg, Phys. Rev. {\bf
147}, 295 (1966).

\bibitem{Scharnberg80}
K. Scharnberg and R. A. Klemm, Phys. Rev B {\bf 22}, 5233 (1980).

\bibitem{Lukyanchuk87}
I. A. Luk'yanchuk and V. P. Mineev, Sov. Phys. JETP {\bf 66}, 1168
(1987).

\bibitem{Shimahara96}
H. Shimahara, S. Matsuo, and K. Nagai, Phys. Rev. B {\bf 53}, 12284
(1996)

\bibitem{Suginishi06}
Y. Suginishi and H. Shimahara, Phys. Rev. B {\bf 74}, 024518 (2006).

\bibitem{Izawa02}
K. Izawa, H. Yamaguchi, T. Sasaki, and Y. Matsuda, Phys. Rev. Lett.
{\bf 88}, 027002 (2002).




\end{thebibliography}
\end{document}